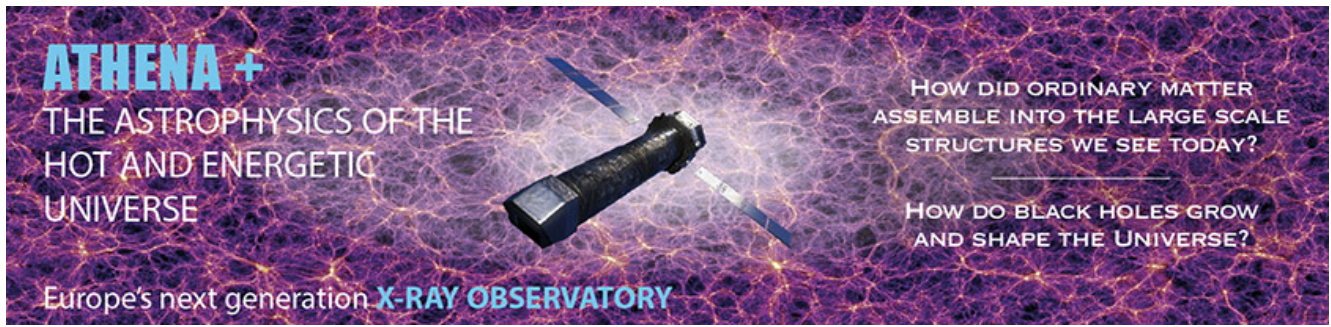

# The Hot and Energetic Universe

An *Athena+* supporting paper

# The Optical Design of the *Athena+* Mirror

Authors and contributors

**Richard Willingale**, Giovanni Pareschi, Finn Christensen and Jan-Willem den Herder

# 1. EXECUTIVE SUMMARY

The manufacture and performance of the *Athena+* X-ray mirror is of paramount importance to the success of the mission. In order to provide a collecting area of ~2 m$^2$ at 1 keV an aperture of diameter ~3 m must be densely populated with grazing incidence X-ray optics. To achieve an angular resolution of ~5 arc seconds these optics must be of extremely high precision and aligned to tight tolerances. A large field of view of ~40 arc minutes diameter is possible using a combination of innovative technology and careful optical design. The Silicon Pore Optics technology (SPO) which will deliver the impressive performance of the *Athena+* primary mirror was developed uniquely by ESA and Cosine Measurement Systems specifically for the next generation of X-ray observatories and *Athena+* represents the culmination of over 10 years of intensive technology development effort. We here describe the X-ray optics design, using SPO, which makes *Athena+* possible.

# 2. INTRODUCTION

The area of the *Athena+* mirror is specified as 2 m$^2$ at 1 keV and the angular resolution as 5 arc seconds Half Energy Width (HEW). This unique combination of large area and high angular resolution provides the ground breaking leap in sensitivity required to achieve the science goals and sets *Athena+* apart from all previous X-ray telescopes. This is illustrated in Fig. 1. If the area at 1 keV is less than ~0.1 m$^2$ then most cosmic sources are photon starved and the observation times required are prohibitively long. If the HEW is greater than ~10 arc seconds source confusion in deep exposures is unacceptable. *Athena+* is in the Golden Quadrant which provides very high sensitivity and minimal source confusion. In addition the field of view provided by the *Athena+* mirror is specified to be a diameter of 40 arc minutes. The vignetting, loss in effective area towards the edge of the field of view, and degradation of the angular resolution at off-axis angles are minimized so that the grasp (collecting area times solid angle product) is maximized and the sensitivity remains high while confusion is low across the full field of view.

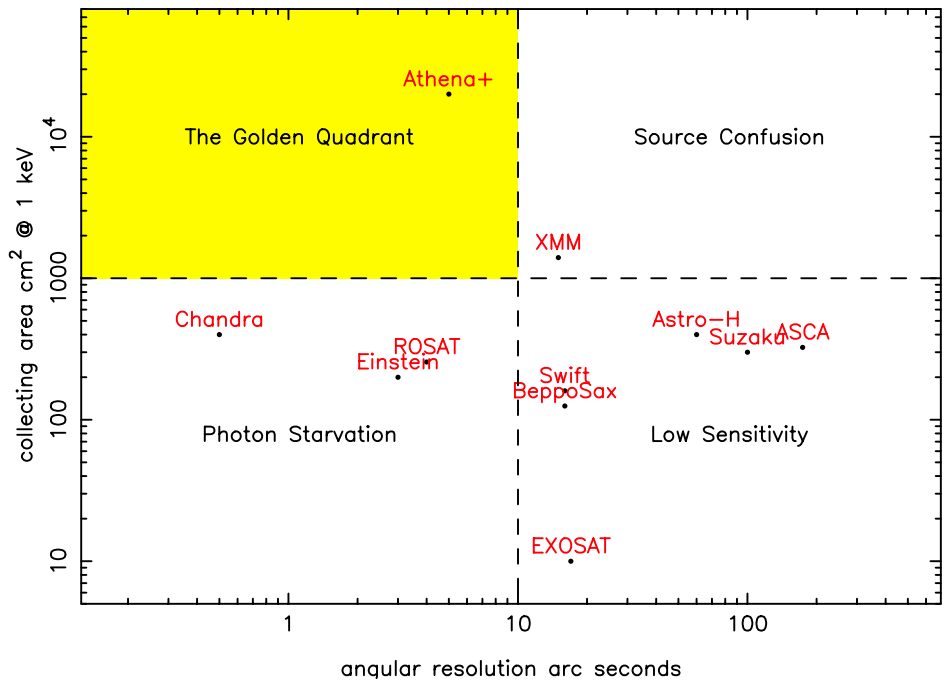

Figure 1: The performance of X-ray telescope modules.

Fig. 2 shows the X-ray mirror technologies available for the construction of X-ray telescopes. The *Athena+* mirror relies on the development of new technology which can provide the large area and high angular resolution within a mass budget and size envelope that is available using the largest launch vehicles. The Silicon pore capability satisfies all these constraints. An alternative, also under development, is segmented slumped glass, Pareschi et al. (2011); Ghigo et al. (2012). This has the potential to provide a similar level of performance but currently has a lower Technology Readiness Level (TRL) than the Silicon pores.

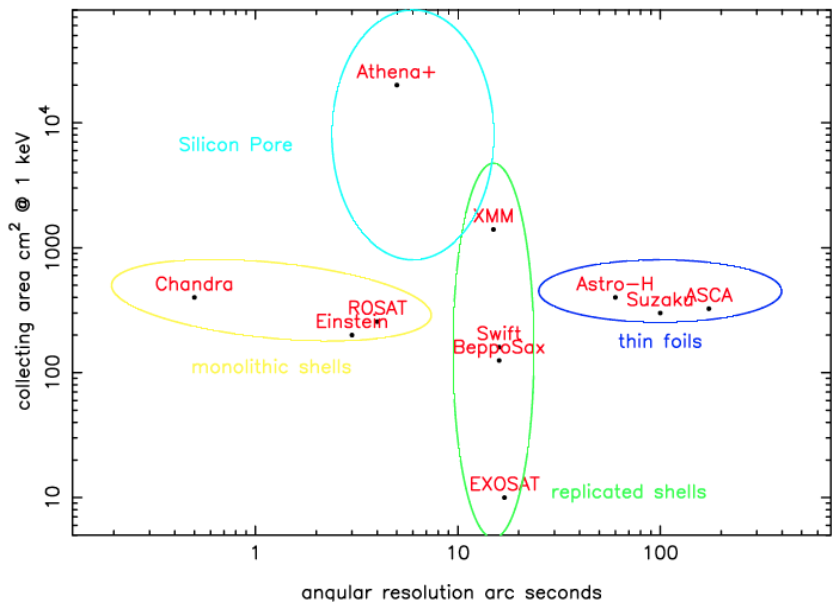

Figure 2: Technology for X-ray telescope modules.

The *Athena+* mirror design is a direct development of the XEUS mirror concept originally described in Aschenbach et al. (2001). It utilises Silicon Pore Optics (SPO) technology first introduced by Beijersbergen et al. (2004) and subsequently applied to the ESA XEUS mission, (Kraft et al., 2005), and later for IXO, (Collon et al., 2010). The SPO technology has now been under development by ESA and Cosine Measurement Systems (http://cosine.nl) for over a decade. SPO utilises commercially available Si wafers which have surface figure and roughness quality ideally suited to X-ray optics applications. The manufacture of Si pores using wafers is illustrated in Fig. 3. The wafers are diced into rectangles typically 60 mm wide and with varying heights.

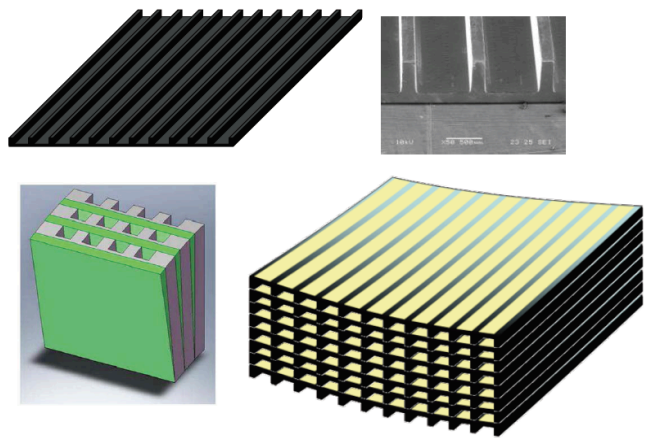

Figure 3: Stacking of Silicon wafers to create a SPO stack. Top left: each Si wafer is diced, wedged and grooves are cut. Top right: groove profile at the end of a Si wafer. Bottom left: the wedge angle between successive plates is achieved by deposition of a very thin precision wedge of material on to the wafers. Bottom right: the reflective coating is applied in strips and the plates are curved and bonded to form a rigid stack.

A thin wedge of material is deposited onto both sides of the wafer so that when the wafers are stacked the reflecting surfaces are arranged in a radial pattern which provides a common in-plane focus. Regular grooves, with a rectangular profile, are cut leaving a thin membrane of thickness ~0.15 mm which supports the entire reflecting surface. The sides of the grooves form parallel ribs which also have a thickness of ~0.15 mm. The faces at the tops of the ribs are untouched and when the wafers are pressed together they cold-bond to the surface of the adjacent wafer, without any gluing, and form a rigid block containing an array of very regular, rectangular pores. The reflecting surfaces are coated with high-Z material (e.g. Iridium or Gold) leaving uncoated strips so that the top of the ribs of one wafer match the pristine Si strips in the next wafer allowing the cold-bond to be made securely. The wafers are curved to the appropriate radius of curvature using a precision mandrel so that reflecting surfaces in all the pores match the surface of revolution required in a Wolter I optical system. This curvature provides the out-of-plane focusing between successive pores across the width of each wafer. The wedge angle and azimuthal curvature of the wafers ensure that all the pores in a given stack point towards a common vertex. A SPO module comprises two wafer stacks. The first stack is a narrow sector of a nest of surfaces of revolution which approximate the paraboloid surfaces (1st reflection) in a nested Wolter I system. Similarly, the second stack provides an approximation to the hyperboloid (2nd reflection) in the nested Wolter I system. The grazing angles of the two reflections are set to be equal so the kink angle between the axis of pores in the 1st stack and the axis of pores in the 2nd stack must be set precisely at twice the grazing angle. Fig. 4 shows a complete module. The mounting plates incorporate 3 mounting lugs and pins which are used as an isostatic mount when integrating the complete SPO modules into the mirror aperture.

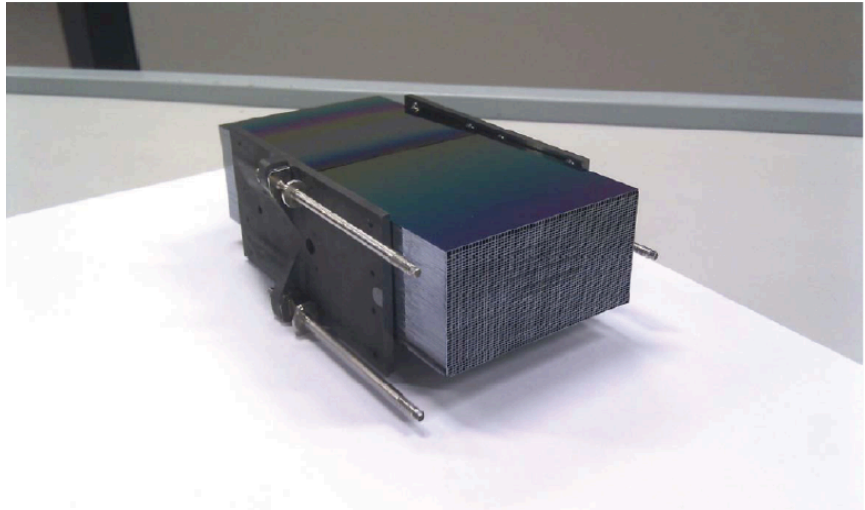

Figure 4: A complete SPO module comprising two stacks. The kink angle between the front and back stack is set accurately and secured by glueing the stacks between the mounting plates.

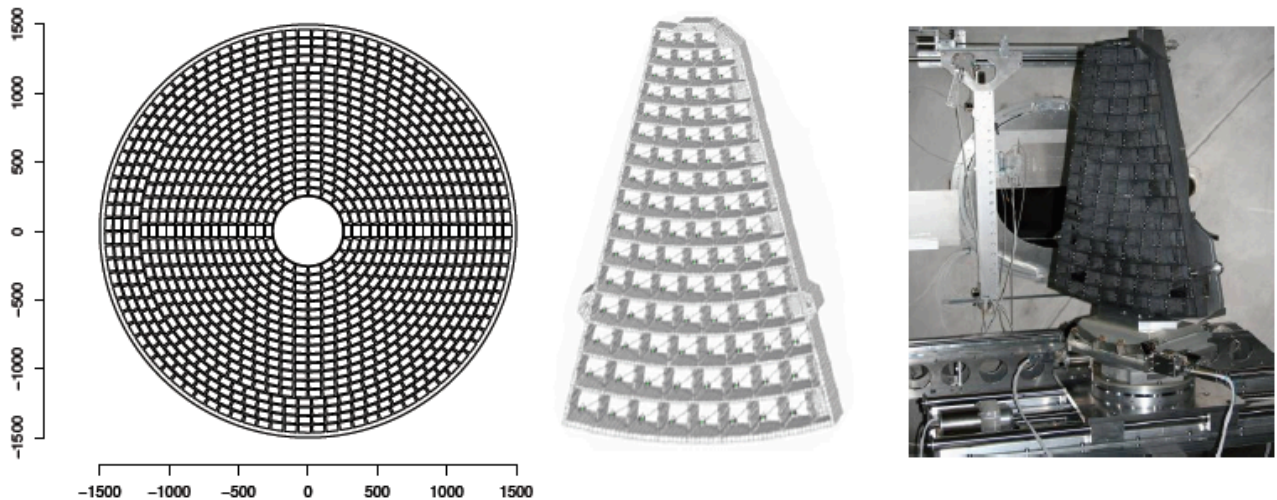

Figure 5: Integration of SPO modules to create the complete mirror. Left: Modules are arranged in rings to populate the aperture. Centre: For mechanical convenience the aperture can be split into sectors. Right: A prototype sector-petal constructed and tested as a proof of concept.

Fig. 5 shows how the full aperture is populated by the individual modules. For simplicity the size of the modules shown in the left-hand diagram is fixed and the azimuthal and radial spacing between the modules has been set conservatively to allow for ample support structure. The centre and right-hand pictures indicate a possible mechanical arrangement using sectors or so-called petals. The size of the modules and the way they are packed can be optimized to ensure that the specified collecting area can be achieved and that the completed mirror system is mechanically robust.

## 3. PORE GEOMETRY

The geometry of a single pore within a SPO module is shown in Fig. 6. The coated reflecting surfaces are shown in red and the double grazing angle reflection of a single ray is indicated in blue. The entrance and exit aperture of the pore are plotted in green. The radial width of the pore (in the plane of reflection) is determined by the standard thickness of the Si wafers, $t_w = 0.775$ mm, and the depth of the cut grooves. In the current production the membrane thickness is set to $w_m = 0.17$ mm so the radial width of the pores is $d = 0.605$ mm.

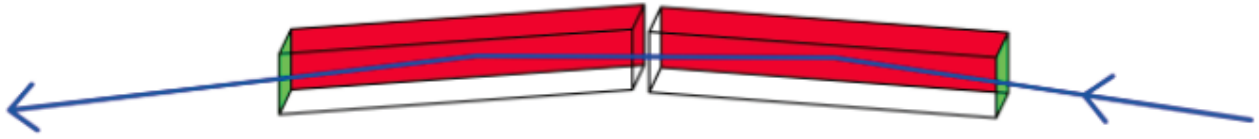

Figure 6: Single pore geometry within a module. The pore has two sections corresponding to the parabolic and hyperbolic surfaces of the Wolter I. The small gap between the sections contains the principal plane of the mirror system.

In order to maximise the collecting area of the pore for on-axis rays the axial length of each reflecting surface, L, (the height of the diced wafers) must be set such that

$$\frac{d}{L} = \tan(\theta_g) \approx \theta_g \qquad (1)$$

where $\theta_g$ is the grazing angle. The grazing angle is determined by the radial position of the pore in the aperture, R.

$$\frac{R}{F} = \tan(4\theta_g) \qquad (2)$$

where F is the focal length. So for the small grazing angles required to yield a high reflection efficiency for soft X-rays ($\theta_g < 2$ degrees) we have that

$$L \approx \frac{4Fd}{R} \qquad (3)$$

The axial length of each module is proportional to the inverse of the radial position in the aperture. Fig. 7 shows the distribution for the module rings shown in Fig. 5. The inner ring of modules at R=285 mm has an axial length of L = 101.9 mm for each stack. The outer ring at R=1437 mm has L=20.3 mm.

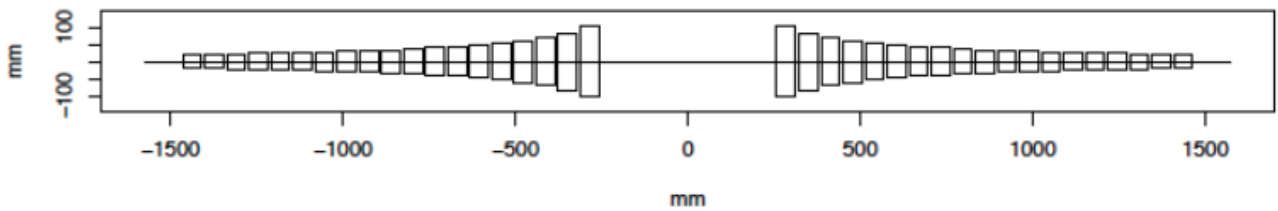

Figure 7: Axial length distribution of SPO modules across the aperture.

If the reflecting surfaces in each pore were planar then there would be no focusing of X-rays within the pore and on-axis rays would form a rectangular patch of size $d = 0.605$ mm in the focal plane. The azimuthal curvature introduced using the mandrel during stacking (see Fig. 3) converts the reflecting surface from a flat plane in to a cone and therefore produces a conical approximation to the Wolter I geometry. This imparts out-of-plane focusing within each pore (and across each module) such that the Point Spread Function (PSF) from a single pore is a line. If there are no figure errors and the reflecting surfaces are perfect cones the length of this line will be $d = 0.605$ mm. Providing all the modules are correctly aligned the line foci from the combination of the millions of pores across the full aperture will overlap to form a PSF with circular symmetry. The idealized form of the PSF for the conical approximation is shown in Fig. 8. The full width of the PSF disk is $d = 0.605$ mm and the HEW is half this. With a focal length F=12 m this HEW corresponds to 5.2 arc seconds. Therefore using the conical approximation with this focal length limits the angular resolution defined by the HEW to 5 arc seconds or worse. If the figure and alignment errors are very small then the PSF will have a sharp central core which provides much better angular resolution which may be useful for bright sources.

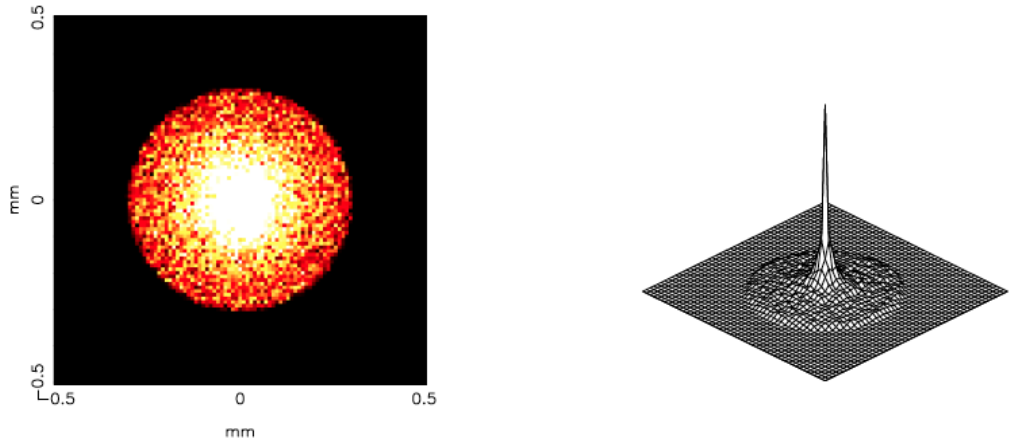

Figure 8: The PSF from the conical approximation with no figure or alignment errors.

In practice in-plane figure errors will limit the size of the PSF. If the conical approximation is used the length of the line focus from each pore will be > 0.605 mm. Fig. 9 shows the PSF from a single module. The line foci from the individual pores is spread out into a bow-tie shape. The measured PSF from a 45 plate stack is shown to the right. In this case stacking errors introduce in-plane figure errors which push the HEW up to 16.6 arc seconds. These errors will be greatly reduced using the latest generation stacking robot and a cleaner stacking environment. The central neck of the measured PSF is broadened because of the finite divergence of the input X-ray beam. Further details about the error terms which contribute to the angular resolution are discussed in Section 6.

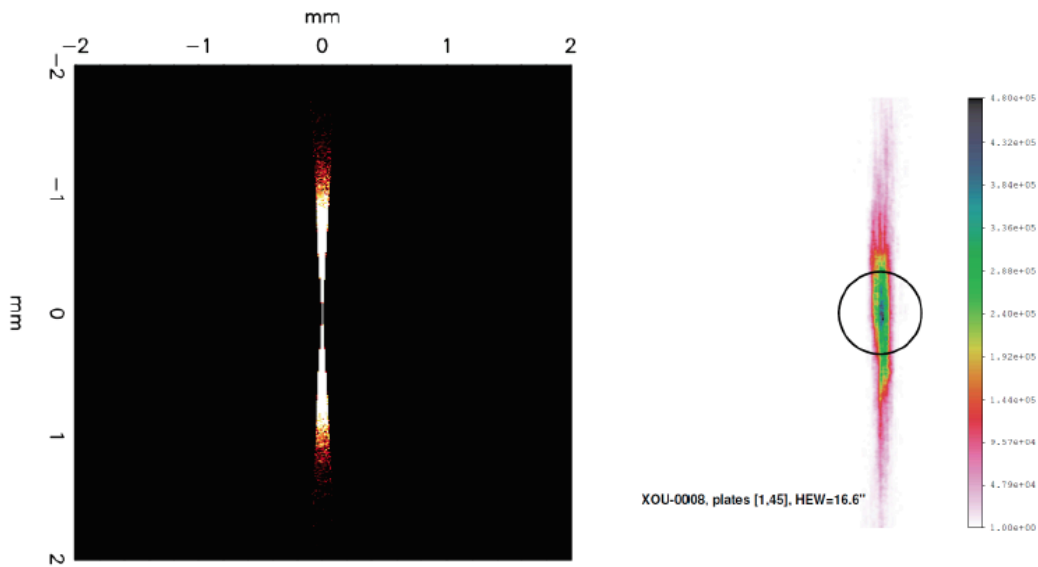

Figure 9: The PSF of a SPO module. Left: ray tracing prediction using the conical approximation and including in-plane figure errors. Right: the PSF of a 45 plate stack measured in X-rays.

## 4. SPECIFYING THE REFLECTING SURFACES FOR THE CONICAL APPROXIMATION SPO MODULES

The conventional origin datum for determining the Wolter I equations is the position of the on-axis focus, van Speybroeck & Chase (1972). A more convenient datum for the SPO surfaces is the intersection of the optical axis with the join plane between the 1st and 2nd surface stacks. If x is the axial coordinate then the 1st surface pores are at positive x and the 2nd surface pores are at negative x. We assume that the grazing angles on the 1st and 2nd surfaces, for on-axis rays, are equal giving maximum throughput, as already discussed above. The ratio of the pore width to pore length, d/L, is given by Equation 1 and for a pore at radius R in the aperture the optimum axial length of the pores is given by equation 3. The grazing angle is given by

$$\theta_g = \frac{1}{4}\arctan\left(\frac{R + d/2}{F}\right) \quad (4)$$

where F is the focal length, the axial distance from the join plane to the on-axis focus. Note that the radius is increased a little by d/2. This ensures that rays which intersect the reflection surface at the centre of the pore are brought to the correct focus at a distance F from the join plane.

The equation for the axial profile of the 1st surface in the conical approximation is $r_1 = \tan(\theta_g)x + R$. Writing this in the more conventional form for the conic section generators we have

$$r_1^2 = \tan^2(\theta_g)x^2 + 2\tan(\theta_g)Rx + R^2 \quad (5)$$

and similarly, we can define the 2nd surface which has a cone angle 3 times that of the 1st surface as

$$r_2^2 = \tan^2(3\theta_g)x^2 + 2\tan(3\theta_g)Rx + R^2 \quad (6)$$

The conical axial profiles defined by Equations 5 and 6 are plotted as the dashed lines in Fig. 10.

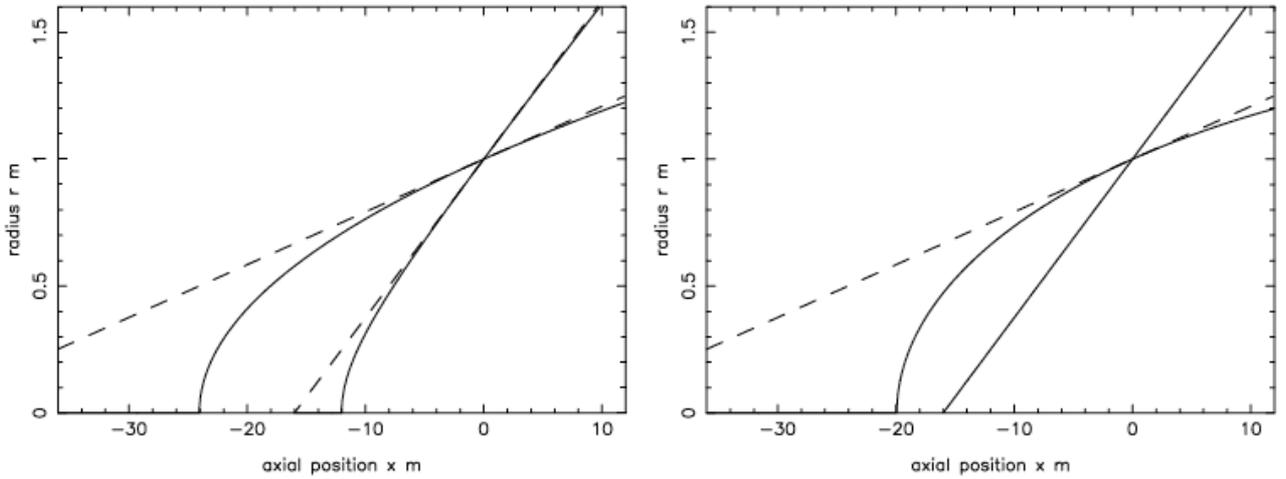

Figure 10: The axial profiles for a pore at radius R = 1 m and focal length F = 12 m. Left-hand panel: The conical approximation axial profiles, dashed lines, and equal curvature axial profiles, solid lines. Right-hand panel: a double curvature 1st surface axial profile with a conical approximation 2nd surface.

Within a SPO stack the surface profiles are controlled by the angle of the wedge deposited onto the wafer surfaces. With a wafer thickness $t_w$ the wedge angles for the 1st and 2nd surfaces are given by

$$\theta_{w1} \approx \frac{t_w}{4F} \quad (7)$$

$$\theta_{w2} \approx \frac{3t_w}{4F} \quad (8)$$

## 5. CORRECTING THE CONICAL APPROXIMATION ABERRATION

In order to improve the angular resolution of the SPO we need to introduce some curvature into the axial profiles of the pores as described in Willingale & Spaan (2010). The full width of the conical approximation PSF shown in Fig. 9 is equal to the radial pore size, d, and to eliminate this spreading we need curvature along the length L of one or both of the 1$^{st}$ and 2$^{nd}$ surfaces. X-rays at the outer edges of the PSF are reflected from close to the ends of the pores and these must be deflected by d/(2F) radians to bring them into focus. This can be achieved by increasing the grazing angles at the ends of the pores by d/(8F) radians. The ends of the pores are an axial distance of L/2 from the pore centre so the axial curvature required is given by

$$\frac{d^2r}{dx^2} = -\frac{d}{4FL} = -\frac{\tan(\theta_g)}{4F} \approx -\frac{R}{16F^2} \qquad (9)$$

This curvature can be included using a extra term $\alpha(x-L/2)^2$ in Equation 5 and $\alpha(x + L/2)^2$ in Equation 6 where

$$\alpha = R\frac{d^2r}{dx^2} = -\frac{R\tan(\theta_g)}{4F} = -\tan^2(\theta_g) \qquad (10)$$

The presence of the offset axial positions x/L−2 and x+L/2 ensures that the gradients at the centre of the pore axial profiles are the same as in the original conical approximation. The axial profiles are then

$$r_1^2 = \tan^2(\theta_g)x^2 + 2\tan(\theta_g)Rx + R^2 - \tan^2(\theta_g)(x - L/2)^2 \qquad (11)$$

$$r_2^2 = \tan^2(3\theta_g)x^2 + 2\tan(3\theta_g)Rx + R^2 - \tan^2(\theta_g)(x + L/2)^2 \qquad (12)$$

Equations 11 and 12 are the *equal curvature axial profiles* plotted as the solid curves in the left-hand panel of Fig. 10. These profiles are not the same as the true Wolter I parabola-hyperbola but they are very close, particularly in the vicinity of the join plane. Ray tracing the full *Athena+* aperture using these equal curvature axial profiles for all the pores gives a HEW of ≈0.1 arc seconds on-axis with a very small /0.2 mm axial shift of the optimum focal position.

It is not necessary to impose curvature on both the 1$^{st}$ and 2$^{nd}$ reflecting surfaces. We can, for example, use double the curvature for the 1$^{st}$ surface using $\alpha = -2\tan^2(\theta_g)$ and leave the 2$^{nd}$ surface as a simple cone. This configuration is shown in the right-hand panel of Fig. 10 and, remarkably, it gives an on-axis performance almost identical to the *equal curvature profiles*.

The optimum axial curvature required, as described above, gives a sagittal deviation of $\Delta r = dL/32F = d^2/(8R)$. With the baseline radial pore size of d = 0.605 mm, $\Delta r$ = 0.16 microns for the inner modules at R = 0.285 m with L = 101.9 mm and $\Delta r$ = 0.03 microns for the outer stacks at R = 1.437 m with L = 20.3 mm. These sag values are equivalent to a constant axial slope change of $\Delta\theta = d/(16F)$ at the ends of the pores compared to the conical approximation design across all the pores in the aperture. Using the baseline pore size and focal length this is 0.65 arc seconds. i.e. the $\Delta r$ sag values or $\Delta\theta$ gradient values required are very small for all positions in the aperture and for all the axial stack lengths required.

## 6. VIGNETTING

The vignetting function is determined by collimation imposed by the pore geometry in the SPO modules. In essence each SPO module acts as a rectangular pore collimator but with the added complications. The outer walls of the pores are highly reflecting while the remaining 3 walls are rough and absorb X-rays effectively and there is a kink angle half way down the pore which divides the reflecting wall into the two Wolter I reflecting surfaces. The radial width of the pores is determined by the initial wafer thickness and fixed at around d = 0.605 mm. The azimuthal width of the pores is determined by the rib spacing which is set at $d_{rib} \approx 1$ mm in the current SPO production. If the tooling used to cut the grooves is changed the rib spacing can be increased. It is expected that $d_{rib} \approx 3$ mm is possible without significant loss in the mechanical integrity of the completed stack.

For a source on-axis all the pores reflect in a plane containing the radius vector with 2 reflections bringing all the flux to a focus. We will refer to this as the in-plane direction. This is illustrated by the left-hand schematic in Fig. 11.

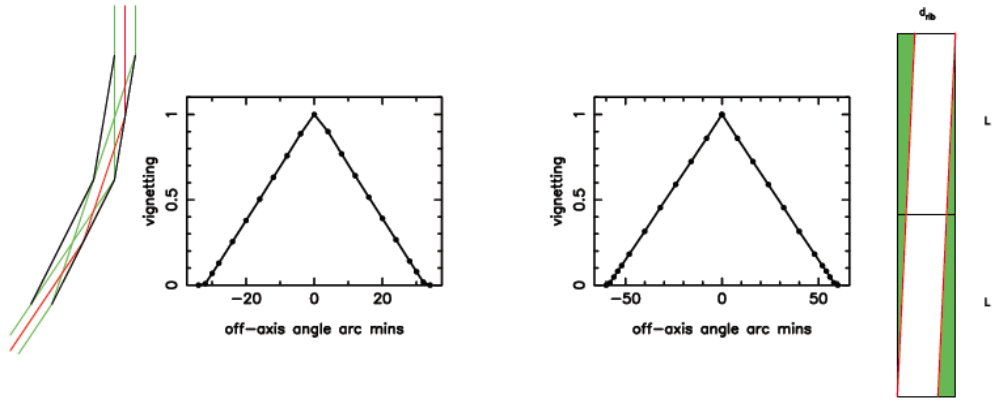

Figure 11: Left-hand schematic: in-plane rays on-axis. Green rays are the extremes with separation d. The central ray is shown in red. Left-hand graph: the in-plane vignetting function for a module at radius R = 850 mm, axial length L = 32 mm at 1.25 keV. Right-hand graph: the out-of-plane vignetting function for the same module and energy, drib = 1 mm. Right-hand schematic: the loss of area for an out-of-plane direction.

The graphs in Fig. 11 show the vignetting function of a mid-aperture module at R = 850 mm with axial length L = 32 mm and $d_{rib}$ = 1 mm. We expect the out-of-plane profiles to be symmetrical about the origin and this is indeed the case. The in-plane profiles show a little asymmetry because the grazing reflection angles are different for positive and negative off-axis angles. However at 1.25 keV this effect is minor and both the in-plane and out-of-plane profiles are closely symmetrical and driven by simple geometric shadowing giving the characteristic triangular shapes as plotted.

The geometry of the out-of-plane vignetting is illustrated by the right-hand schematic in Fig. 11. The green areas of the reflecting surfaces are lost as a source moves off-axis. For a source on-axis the geometric collecting area of the pore is $Ld_{rib}$ but for an out-of-plane off-axis angle β the geometric collecting area of a single pore falls as

$$A(\beta) = Ld_{rib} - 2L^2 |\tan \beta| \qquad (13)$$

where $d_{rib}$ = 1 mm is the pore size in the azimuthal direction on the aperture. We expect this area to drop to zero at β≈$d_{rib}$/2L. Putting in the parameters gives 54 arc minutes in agreement with the ray tracing results. The geometry of the in-plane profile is a more tricky to analyse but empirically we find the area drops to zero at an in-plane off-axis angle γ≈ d/2L which is ~32 arc minutes close to the result from the ray tracing.

For a given off-axis position modules across the aperture as shown in Fig. 5 will suffer different combinations of in-plane and out-of-plane vignetting. By ray tracing through all the modules we can derive the vignetting function of the complete mirror system. Fig. 12 shows the vignetting function obtained using different $d_{rib}$ spacing values. Note that as the rib spacing increases the open area increases and provides a higher effective area on-axis.

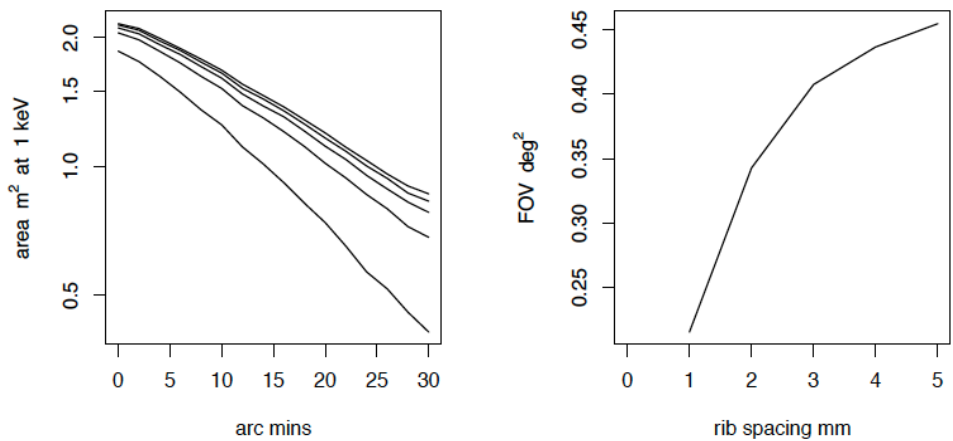

Figure 12: Left-hand panel: the vignetting at 1 keV for drib = 1 mm (lowest curve) through 2,3,4 and 5 mm (highest curve). Right-hand panel: the effective field of view area as a function of drib.

We can define the effective field of view as the sky area for which the collecting area is greater than half the on-axis value. This is plotted in the right-hand panel of Fig. 12. Clearly a rib spacing of ~3 mm provides a significantly larger effective field of view than a spacing of ~1 mm. We adopt $d_{rib}$ = 3 mm as a reasonable tradeoff between on-axis area and field of view against mechanical integrity and a potential increase in figure and alignment errors.

The vignetting function is strongly dependant on the photon energy as illustrated in Fig. 13. Because both the on-axis area and size of the effective field of view decrease with increasing photon energy the grasp for a particular detector size is very energy dependant as shown in the right-panel of Fig. 13. For a FOV 40 arc minutes in diameter the grasp is ~0.5 $m^2$ $deg^2$ at 1 keV.

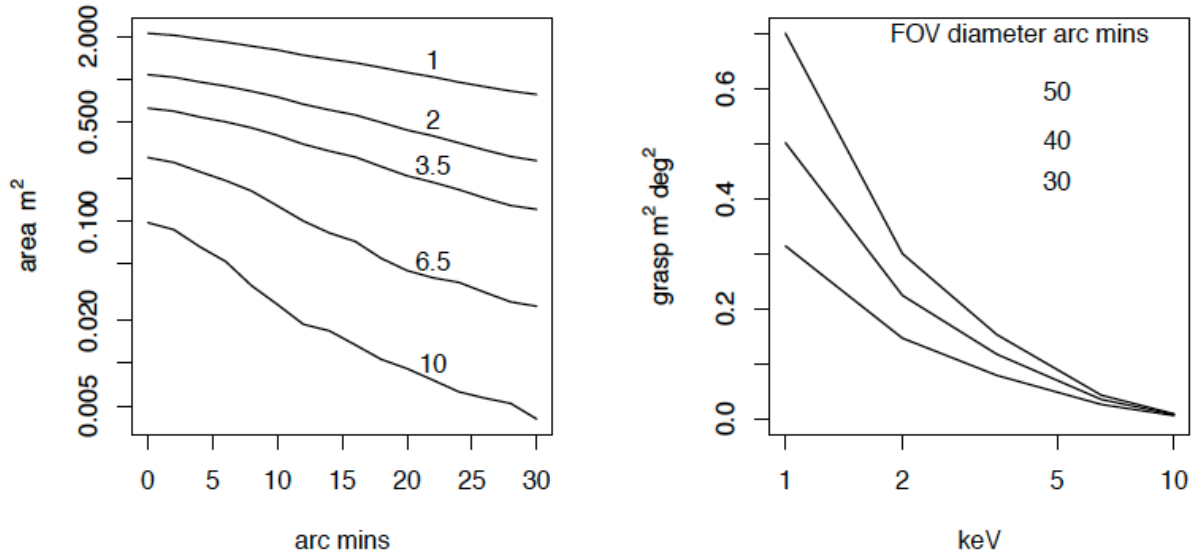

Figure 13: Left-hand panel: the vignetting for drib = 3 mm at 1, 2, 3.5, 6.5 and 10.0 keV. Right-hand panel: the grasp as a function of energy for 3 fields of view as indicated.

## 7. ERROR BUDGET FOR THE ANGULAR RESOLUTION

The following sources of error contribute to the size and quality of the PSF and hence the angular resolution.

- In-plane figure errors. These are responsible for the increase in the length of the module line focus as illustrated in Fig. 9. Curvature errors like the conical approximation to the Wolter I profiles effectively increase this error term. Such errors are most conveniently expressed as a gradient error. A typical allocation for the random contribution is 1 arc sec rms.
- Out-of-plane figure errors. These arise from circularity errors in the stack or rotational misalignment between the 1st and 2nd stacks in the module. A typical allocation is the same order as for the in-plane errors. Such errors increase the width of the line focus but at grazing incidence are diminished by a factor ~$\theta_g$ so are expected to be relatively unimportant.
- Focal length errors. These can arise from a combination of other errors within the module including kink angle errors and wedge errors. If small they are equivalent to and can be compensated by an axial shift of the module. A typical allocation is 1.0 mm.
- Module rotation about the optical axis. Such errors have to be controlled when integrating the modules into the support structure. A rotation $\Delta\theta$ shifts the line focus of a module sideways by a distance $R_{mod}\Delta\theta$. Therefore modules at the edge of the aperture at large Rmod are much more sensitive to these errors. A typical allocation is 2 arc sec rms.
- Module shift in the aperture plane. These errors translate directly to a shift of the module line focus in the focal plane. A typical allocation is 0.05 mm rms.
- Module tilt errors. Because the modules act as individual lenslets the PSF is insensitive to tilts. If a module tilts the centroid of the PSF remains stationary but gets slightly broader because the tilt is effectively a shift to an off-axis angle. The integration process is expected to be very insensitive to these errors. If a stray X-ray grid baffle is employed large tilts may compromise the vignetting function. A typical allocation is 1 arc minute.

# 8. ANGULAR RESOLUTION OFF-AXIS

The angular resolution of a Wolter I optic degrades rapidly with off-axis angle. The situation can be improved using the Wolter-Schwartzschild (W-S) design using surfaces of revolution which exactly fulfill the Abbe sine condition, Chase & van Speybroeck (1973). In a nest of shells of the W-S design the join plane (principal surface of the optic) must necessarily be a sphere of radius equal to the focal length rather than a flat plane as in the conventional Wolter I design. W-S telescopes give a better off-axis performance, in particular when the grazing angles are large, >1.5 degrees and they have been used successfully in the EUV, for example for the ROSAT Wide Field Camera mirrors, Willingale (1988). The off-axis PSF can also be improved using surface figures based on polynomia rather than the conventional parabola-hyperbola used for the Wolter I. Such surfaces have been proposed for the Wide field X-ray Telescope (WFXT), Conconi et al. (2010).

In the *Athena+* design, utilising SPO, the axial lengths of the reflecting surfaces are necessarily short, as shown in Fig. 7. The curvature of the axial profiles can be set to improve the on-axis PSF, as described in Section 4, but there is little scope for further modification of the axial profiles to optimise the off-axis response. However it is possible to produce an approximation to a nested W-S design by placing the join planes of the individual SPO modules on a spherical surface rather than a flat plane. This is illustrated in Fig. 14.

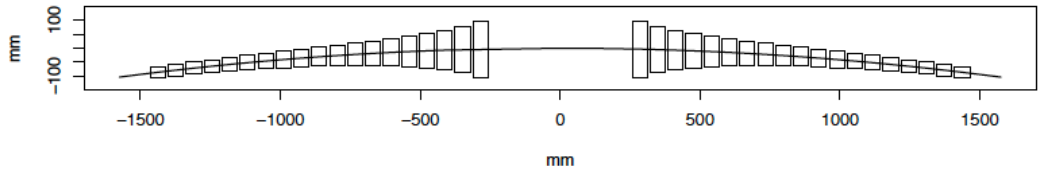

Figure 14: SPO modules integrated across a spherical principal plane in the Wolter-Schwartzschild configuration.

The axial shift required for a modules at radius $R_{mod}$ is $\Delta x = F - \sqrt{(F^2 - R_{mod}^2)}$. If this configuration is used the grazing angles, $\theta_g$ as defined by Equation 4, and therefore the kink angle between the 1st and 2nd surfaces, $2\theta_g$, must be set to

$$\theta_g = \frac{1}{4}\arctan\left(\frac{R + d/2}{F - \Delta x}\right) \qquad (14)$$

A summary of the off-axis HEW that can be achieved using the Wolter I and W-S geometry is shown in Fig. 15.

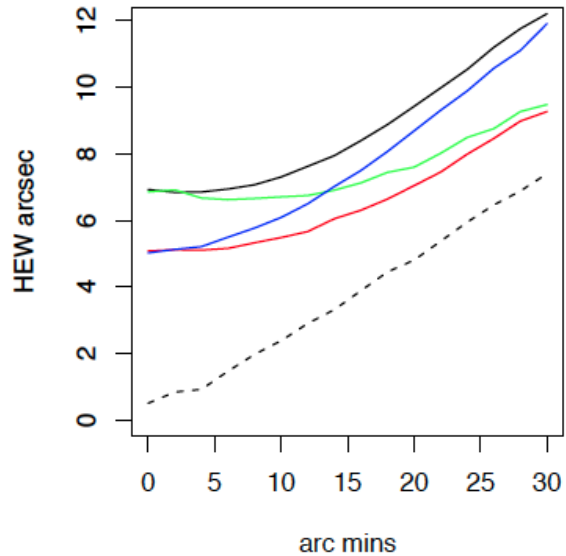

Figure 15: Off-axis angular resolution, HEW. Black - conical approximation with flat principal plane. Green - conical approximation with spherical principal plane. Blue - optimum axial curvature with flat principal plane. Red - optimum axial curvature with spherical principal plane. All these curves include identical figure and alignment errors as described in Section 6. The dashed curve is the limiting angular resolution which can be achieved using the W-S spherical principal plane and no figure of alignment errors.

The W-S configuration (spherical principal plane) offers a significant advantage over Wolter I (at principal plane) because of the improved off-axis response. The rms HEW (averaged over the field of view) is 6.0 arc seconds for the W-S with axial curvature and a FOV diameter of 40 arc minutes. This increases to 6.5 arc seconds if the FOV has a diameter of 50 arc minutes. Using the Wolter I configuration the equivalent rms values are 7.9 and 8.6 arc seconds. Using the conical approximation increases these rms HEW values by ~2 arc seconds but the W-S still offers a significant advantage.

## 9. STRAY X-RAYS

Figure 16 shows the distribution of X-ray flux at 1 keV over the focal plane from a source at 30 arc minutes off-axis.

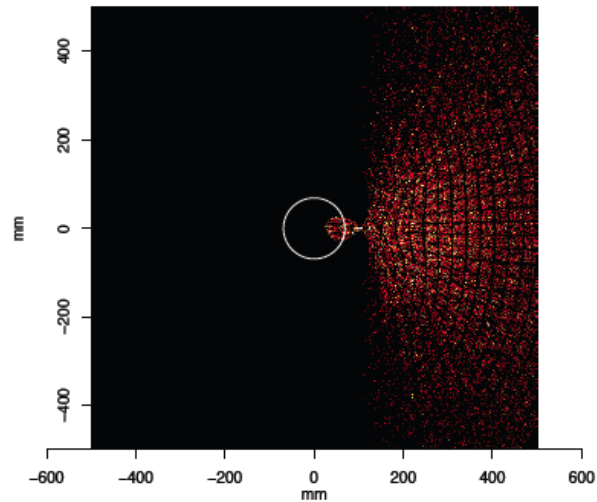

Figure 16: Stray X-rays from single reflections in the pores. 1 keV flux over the focal plane for a source 30 arc minutes off-axis. The imaged (2-reflections) point source is at the neck of the distribution 105 mm from the centre. The circle represents an active field of view diameter 40 arc minutes.

The central circle indicates an active field of view of radius 20 arc minutes. The focused 2-reflection image of the point source is situated at the neck of the distribution, 105 mm from the centre. The rest of the flux is single reflection rays which missed the 1$^{st}$ reflecting surface but intersected the 2$^{nd}$. A small percentage of this stray flux falls on the active field of view. The grid visible in the stray flux pattern arises from the modules distribution over the aperture (Fig. 5). The modules which reflect stray X-rays into the active field of view are situated in-plane with respect to the off-axis source position and have grazing angles close to the off-axis angle. Fig. 17 indicates the geometry of single reflection rays which reach the active field of view. They enter the 1$^{st}$ section of the pore nearly parallel to the reflecting surface and after reflection from the 2$^{nd}$ surface exit the 2$^{nd}$ section in a similar direction to the double reflection imaged rays.

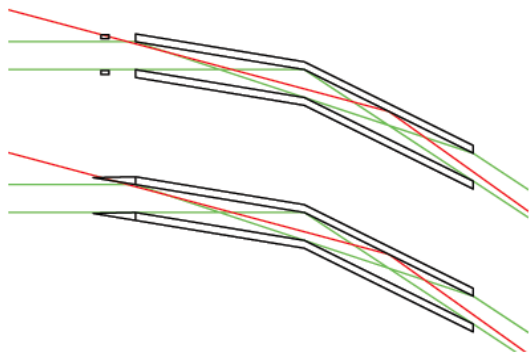

Figure 17: Stray X-rays from single reflections in the pores. Green rays are on-axis, 2-reflection, rays. The red rays miss the 1st surface but reflect from the 2nd and subsequently intersect the active field of view. The top pore shows how a grid placed in front of the pore apertures can block single reflection rays. The bottom pore shows how tapered extensions to the wafers can perform a similar function.

The effective area associated with the stray flux that enters the active field of view has been estimated by ray tracing for a range of off-axis angles. The result is shown in left-hand panel of Figure 18. The distributions are a little ragged because only a small fraction of the rays traced (several million) reached the active detector area thus limiting the statistics. For a given SPO module the peak of the off-axis stray flux occurs close to the grazing angle. The grazing angles range from 18 arc minutes for the inner most pores to 106 arc minutes at the edge of the aperture. The peak of the stray area distribution occurs for a source ~35 arc minutes off-axis. The central region of the field of view, radius 0-5 arc minutes, contains almost no stray X-ray flux (the lowest dashed curve in Fig. 18) while the outer region radii 15-20 arc minutes contains almost half the total (the highest dashed curve in Fig. 18).

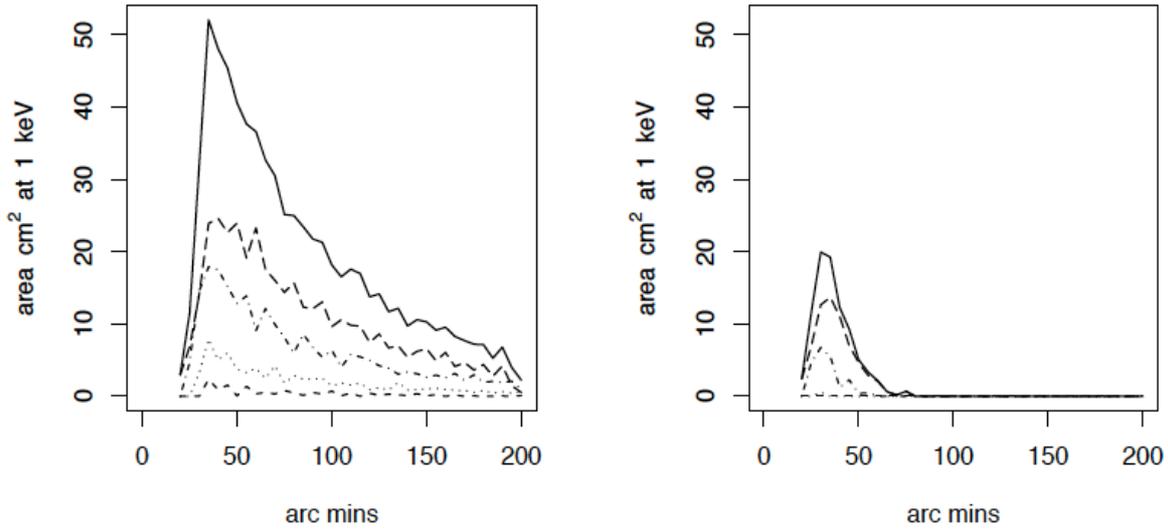

Figure 18: Left-hand panel: The aperture area associated with stray 1 keV X-rays in a 40 arc minutes diameter field of view as a function of off-axis angle. The dashed lines show the area for 4 constituent annuli of the field of view, radii ranges 0-5, 5-10, 10-15 and 15-20 arc minutes. Right-hand panel: The same area curves obtained when a stray X-ray baffle grid is included.

The peak of the stray X-ray area at 1 keV is only ~50 cm$^2$ whereas the on-axis area of the full telescope at 1 keV is ~2 m$^2$ so the ratio of the stray X-ray flux to on-axis area for a particular point source strength is $2.5 \times 10^{-3}$. This sets an upper bound for source contamination introduced by stray X-rays from strong point sources near the active field of view.

The area associated with the stray X-rays is the same for all azimuth angles because of the even distribution of modules across the aperture and the differential stray collecting area-sky area product is $2\pi\beta A(\beta)d\beta$ m$^2$ deg$^2$ where $\beta$ is the off-axis angle in degrees and $A(\beta)$ is the stray area in m$^2$. If we integrate this we get a total stray X-ray grasp at 1 keV of $5 \times 10^{-2}$ m$^2$ deg$^2$. If we multiply this by a uniform diffuse background rate we get an estimate of the stray X-ray rate. The grasp for the imaged X-ray flux is shown in Fig. 13 and for a field of view 40 arc minutes in diameter is ≈0.5 m$^2$ deg$^2$. So the predicted stray X-ray flux from the diffuse background is ~10% of the imaged flux from the same background. Even without any dedicated baffles the extra loading from stray X-rays is modest. For example, the equivalent value for the XMM mirrors without baffles was ~40%.

Two schemes for introducing stray X-ray baffles have been considered by ESA in collaboration with Cosine. These are illustrated in Fig. 17. The first involves extension of the 1$^{st}$ reflection wafers with a tapering profile which will not effect the vignetting of the imaged flux but blocks some of the off-axis rays which miss the 1$^{st}$ reflecting surface but intersect the 2$^{nd}$. Unfortunately there is a limit to how thin the taper can be manufactured (at present a minimum thickness of ~0.1 mm). Such baffles are expected to work reasonably well for the outer modules but not for the inner modules.

The second scheme is a collimating grid mounted in front of the pores. This works in a similar way to the baffle sieve plates used on XMM. In principle they can be very effective but the blocking elements must be a little thinner than the wall thickness of the pores and manufactured and aligned to a very high accuracy so that they don't block on-axis flux and thereby reduce the effective area for the imaged flux. Such a grid has been successfully manufactured and mounted on a module by Cosine.

The efficacy of a grid baffle has been tested by ray tracing. The grid bars were assumed to have thickness 0.15 mm, slightly less than the pore wall thickness $w_m = 0.17$ mm, and a depth equal to the wafer thickness, $t_w = 0.775$ mm. In order to block the stray X-rays the grids must be mounted at a distance $h_g \approx L w_m/d$ along the optical axis of the telescope in front of each module. The ratio $w_m/d = 0.281$ so the grids are mounted just over 1/4 of the axial surface length in front of the modules. It was demonstrated by ray tracing that such grids accurately mounted have no effect on the vignetting function for a field of view 40 arc minutes diameter. Of course any misalignment or change in the grid bar thickness could potentially block some fraction of the imaged flux. The right-hand panel of Fig. 18 shows the area associated with the stray X-rays with the grids in place. The peak of the distribution still occurs for a source at ~35 arc minutes off-axis but is reduced to ~20 cm². So the stray to source flux ratio has been reduced to $10^{-3}$. All the stray flux from off-axis sources > 80 arc minutes has been eliminated and the grasp for stray X-ray flux is of $2.9 \times 10^{-3}$ m² deg². Therefore the stray X-ray flux from a uniform diffuse background has been reduced to ~0.6% of the imaged background.

## 10. AREA AS A FUNCTION OF PHOTON ENERGY

The area as a function of photon energy depends on the high-Z coating used for the reflecting surfaces within the pores. The maximum energy which can be reflected by a module is determined by the grazing angles and hence the radius of the module in the aperture $R_{mod}$ (Equation 4). Each ray suffers two reflections so the collecting area depends on the square of the reflectivity. The left-hand panel of Fig. 19 shows the reflectivity squared as a function of energy for modules at different radii using Iridium with a thin $B_4C$ overcoat. The response of the inner modules extends to above 10 keV while the outer modules only contribute for energies <2 keV. The right-hand panel of Fig. 19 shows the on-axis area vs. energy for two candidate coatings, Iridium and Iridium with an overcoat of $B_4C$. For both coatings the absorption edges of Iridium, in particular the M-edges at 2-3 keV, introduce significant dips in the efficiency. The use or a Carbon based over-coating layer was first suggested by Pareschi et al. (2004) and has the effect of enhancing the low energy reflectivity and filling in the absorption edges to some extent.

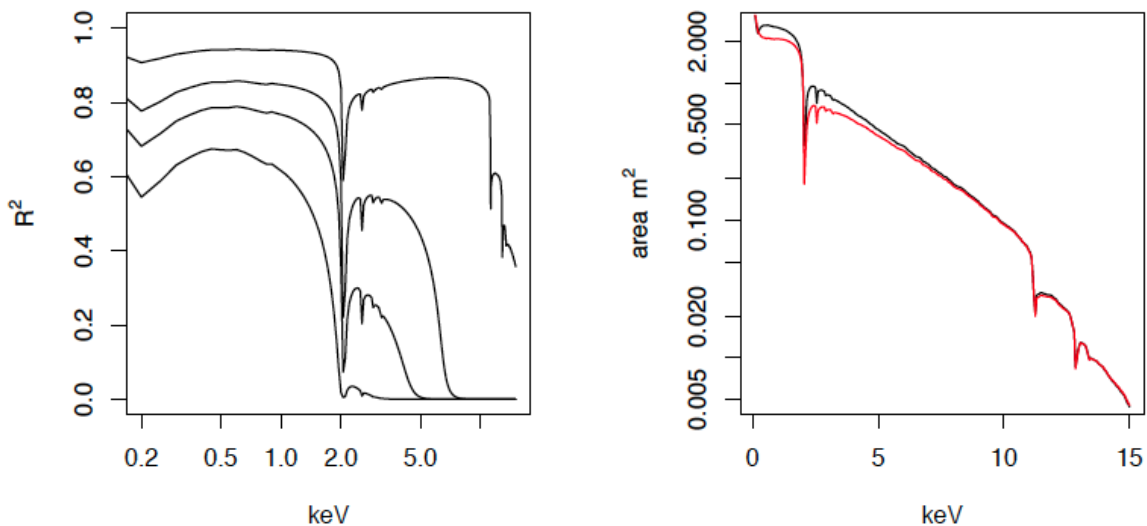

Figure 19: Left-hand panel: reflectivity squared vs. photon energy for the outer modules (lowest curve $\theta_g \approx 18$ arc mins), inner modules (highest curve $\theta_g \approx 106$ arc mins), and for two intermediate module rings ($\theta_g \approx 46$ and 69 arc mins). All curves using Iridium with a B4C overcoat. Right-hand panel: on-axis area vs. energy. Red - Iridium coating. Black - Iridium with B4C overcoat.

More complicated multilayer coatings can be considered to further improve the high energy response (Jakobsen et al., 2011).

## 11. A RAY TRACING MODEL

All the simulation results presented here were produced using a comprehensive ray tracing model. The software runs under the statistical computing package R (R Core Team, 2013) which is freely available (The R Project for Statistical

Computing http://www.r-project.org). The R modules and dynamically loadable shared objects required, and a comprehensive manual, are available from the author (R. Willingale zrw@le.ac.uk). The model incorporates all pertinent aspects of the optics. The physical properties of all the SPO modules are specified individually including dimensions, positions, error budget terms etc. Within each module the pores are specified individually such that distortions and imperfections inherent in the manufacturing process can be modeled accurately. Different reflecting coatings, surface roughness and figure curvature (Wolter I, conical approximation, W-S etc.) can be selected or specified.